# Testing of Solar2000 EUV flux model between 900-1350 Å using Greenline Dayglow Emission

M.V. Sunil Krishna and Vir Singh [†]

Department of Physics, Indian Institute of Technology Roorkee

Roorkee – 247 667, India

[†]e-mail: virphfph@iitr.ernet.in

**Abstract**

The contribution of photodissociation of molecular oxygen to the total volume emission rate of greenline dayglow emission at 5577 Å is modelled in the present study. The Solar EUV radiation fluxes for the modelling are obtained from the Solar2000 V 2.25 model. The modelling has been done in the altitude range of 92 - 105 Km, where the photodissociation and the three body recombination are the main contributing processes to the greenline dayglow emission. The present results are discussed in the light of Wind Imaging Interferometer (WINDII) observations of greenline dayglow emission in the altitude range of 92 - 105 Km. It is found that the Solar2000 V2.25 flux model gives better agreement with the WINDII observations especially in the 92 - 96 Km where the earlier models predicted a very low emission rate. In the mesospheric emission peak region the present results are higher than the measurements and this discrepancy may possibly be attributed due to higher values of solar fluxes for those spectral lines which are main sources of the production of greenline dayglow emission. The present study suggests the reexamination of Solar2000 flux model at least for those spectral lines which are the main sources of greenline dayglow emission in mesospheric emission peak region.

# 1. Introduction

Airglow emissions are found to be a valuable source of information about the composition, dynamics and the chemical state of thermosphere and upper mesosphere. Among the various airglow emissions the atomic oxygen greenline dayglow emission at 5577 Å[1] wavelength has been given a particular attention due to the fact that it is the most readily observed and the brightest emission. In the recent years a lot of attention has been paid to the modelling of this emission[2]-[5].

The consistency of the model depends on various factors such as cross sections, reaction rate coefficients, neutral atmosphere and Solar Extreme Ultra Violet (EUV) radiation fluxes. Though the emission rate depends on the above parameters but the Solar EUV flux has a strong influence on the airglow emissions. Consequently there is a need of more appropriate solar EUV fluxes in the model calculations, so that one can have a better comparison with the observed emission rates. The earlier model studies[2],[6] have used solar fluxes from Hinteregger et al.[7] however these model studies could not reproduce the Wind Imaging Interferometer (WINDII)[8] measurements of greenline dayglow emission between 92 and 105 Km altitude region . The reason for this discrepancy has been attributed to the solar flux models[7],[9] and inadequacies of atomic oxygen density[10] in MSIS-90 neutral model atmosphere. It would be important to mention here that the Solar2000 V2.25 [11] EUV flux model is the latest and has not been used in the earlier model studies. Therefore one should include the Solar 2000 EUV flux model in the earlier model[2] to test the consistency of the Solar EUV fluxes in relation to greenline day glow emission. It would be pertinent to mention here that we would try to identify those regions where Solar-2000 EUV flux model explains the WINDII measurements. Further, in those regions where Solar 2000 EUV flux model does not explain the measurements we would try to find out the inadequacies in Solar2000 EUV flux model.

Singh et. al[2] and Tyagi and Singh [6] have studied greenline dayglow emission at various latitudes and longitudes. The model studies of Singh et. al[2] are based on the photoelectron flux model as developed by Richards and Torr[12] which used Hinteregger et al.[7] solar flux model. Singh et.al [2] have used a linear proxy in scaling the solar flux to account for the solar flux variability. On the other hand Tyagi and Singh [6] used Glow model as developed by Soloman model[13]. In Glow model Hinteregger et al.[7] solar flux was also used but with different scaling procedure to account for the solar flux variability. This scaling procedure has been discussed by Tyagi and Singh [6] and would not be repeated here. Due to different scaling procedures in the above models, the solar fluxes were found quite different particularly in 900-1350Å wavelength region. Consequently none of the model could explain the WINDII measurements between 92 and 105 Km altitude region. We would mention here that the model developed by Singh et. al[2] is quite easy to handle and provides very reasonable results. The problem in this model is the scaling of solar fluxes. Consequently it would be desirable if one should replace the Hinteregger et al.[7] solar fluxes in the model of Singh et. al[2] by Solar2000 EUV fluxes and restudy the greenline dayglow emissions.

In the present study we have included Solar2000 V 2.25 EUV flux model in the modified model of Singh et al[2]. The photodissociation of molecular oxygen which is the main contributing source between 92 and 105 Km for the production of greenline dayglow emission is studied using Solar2000 EUV flux model. The present results are compared with the WINDII observations [8].

## 2 Model

The greenline dayglow emission mainly occurs in lower thermosphere and upper mesosphere. Out of the two regions of occurrence of this emission our concentration in the present study is in the upper mesosphere region (92-105 Km). In this altitude region photodissociation of molecular oxygen by solar EUV photons and three body recombination processes are the main source of greenline dayglow emission as suggested by Wallace and McElroy[14]. The following process gives the production of $O(^1S)$ due to photodissociation of $O_2$.

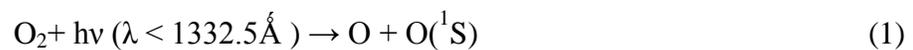

$$O_2 + h\nu \ (\lambda < 1332.5 Å) \rightarrow O + O(^1S) \qquad (1)$$

Three body recombination reaction will also make significant contribution to the total volume emission rate between 95 and 105 Km. This reaction was initially proposed by Chapman[15] which proceeds as follows.

$$O(^3P) + O(^3P) + O(^3P) \rightarrow O_2 + O(^1S) \qquad (2)$$

However it is generally accepted that the production of O($^1$S) is by a two step mechanism as proposed by Barth[16]. This reaction proceeds with the presence of a third body M as

$$O(^3P) + O(^3P) + M \rightarrow O_2^* + M \qquad (3)$$
$$O_2^* + O(^3P) \rightarrow O_2 + O(^1S) \qquad (4)$$

Here $O_2^*$ represents unidentified excited state of molecular oxygen. In the present model, the production of O($^1$S) due to three body recombination process is studied by using the parameters of McDade et al.[17]. The O($^1$S) production rate due to photodissociation of $O_2$ is calculated by using the following equation.

$$P[O(^1S)] = [O_2] \int F_{z,\lambda} Q_\lambda \sigma_\lambda d\lambda \qquad (5)$$

where $F_{z,\lambda}$ is the solar flux at altitude z for the wavelength λ, $Q_\lambda$ is the quantum yield at a wavelength of λ and $\sigma_\lambda$ is the photoabsorption cross section for $O_2$. The photoabsorption cross section and relevant quantum yields in the present study are taken from Fennelly and Torr[18] for the wavelengths between 900-1025Å and from Lawrence and McEwan[19] for the wavelengths from 1025-1200 λ. In the wavelength range between 1200-1350 Å the O($^1$S) quantum yields are very low. In this range the Lyman α (1216 Å) spectral line is one of the most dominant radiation. The flux of Lyman α is very high and may result in the significant production of O($^1$S) due to dissociation of $O_2$ and $CO_2$ even though quantum yield is too low. Singh et. al[2] have discussed in detail about this source and included in the model. In many cases they found that the greenline emission rate shows a wavy nature which is not found in the measurements. Due to large uncertainties in the quantum yield we have not included Lyman α source in the present study. $F_{z,\lambda}$ is taken from Solar2000 EUV flux model V2.25.

## 3  Results and Discussion

The atomic oxygen greenline dayglow emission profiles are modeled at selected geographic locations as observed by WINDII. In Fig 1 a comparison of emission rate due to photodissociation reaction is made between the present results and the results obtained by Singh et al.[2]. It is noticeable from the Fig 1 that the present results are about 1.6 times higher than the existing model results of Singh et al[2]. This difference is due to the fact that Solar2000 EUV fluxes are significantly higher than the EUV fluxes used by Singh et.al.[2] between the wavelengths 900-1350Å. The results of total volume emission rate (VER) of the present study are shown in Fig 2 alongwith the results of Singh et al.[2] and observed values of WINDII at various geographic locations for six cases. It has already been discussed above that the present study does not include the Lyman α source in the calculation of emission rate. Therefore the wavy nature in emission rate is not found in the present results as has been reported by Singh et .al[2]. It is noticeable from Fig 2 that the present results are in better agreement with WINDII observations between 92-96 Km altitude region. These results show an improvement over the results of Singh et al[2]. Further, it is noticeable from the Fig 2 that in majority of cases the present VER in the peak region is higher than the WINDII measurements. This disagreement suggests that Solar 2000 flux model has some inadequacies in the flux for those spectral lines which contribute to the production of O($^1$S) in the peak region. A close examination of the O($^1$S) production rate[19] shows that the solar radiation at 1025Å and 1037 Å wavelengths is the major source of greenline dayglow emission. Further we have examined that the fluxes at 1025Å and 1037Å wavelengths obtained from the Solar 2000 flux model are higher than the Hinteregger et al.[7] solar fluxes. Consequently one may attribute the higher emission rate in peak region due to higher values of solar flux at 1025Å and 1037Å wavelengths in Solar2000 flux model. At this juncture we would suggest that one should check the sensitivity of O($^1$S) production due to solar radiation at 1025Å and 1037Å wavelengths and accordingly the solar flux in Solar 2000 flux model be modified. We are in the process to examine these effects and the results would be reported in near future.

Upadhyaya and Singh[10] obtained a correction factor to the atomic oxygen density by using the results of Singh et al.[2] to reproduce the observed emission profiles of greenline dayglow emission between 92 and 105 Km. The present study also suggests that the correction factor as given by Upadhyaya and Singh[10] should be reviewed in the light of Solar2000 EUV flux model.

## 4  Conclusions

The present model gives a better agreement with the WINDII measurements of greenline dayglow emission compared to the earlier results between 92-96Kms. The discrepancy between the present results and measurements in the mesospheric emission peak region may be attributed due to higher values of solar flux at 1025Å and 1037Å wavelengths in Solar2000 flux model and needs further study. The present study also suggests that the correction factor given by Upadhyaya and Singh[10] should be reexamined in the light of Solar2000 EUV flux model.

**Caption to Figures:**

**Figure 1**. Comparison of emission rate due to photodissociation process between the present results and Singh et al.[2].

**Figure 2**. Comparison of total volume emission rate profiles of 5577Å emission obtained from the present study with the results of Singh et al.[2] and WINDII observations.

Figure 1

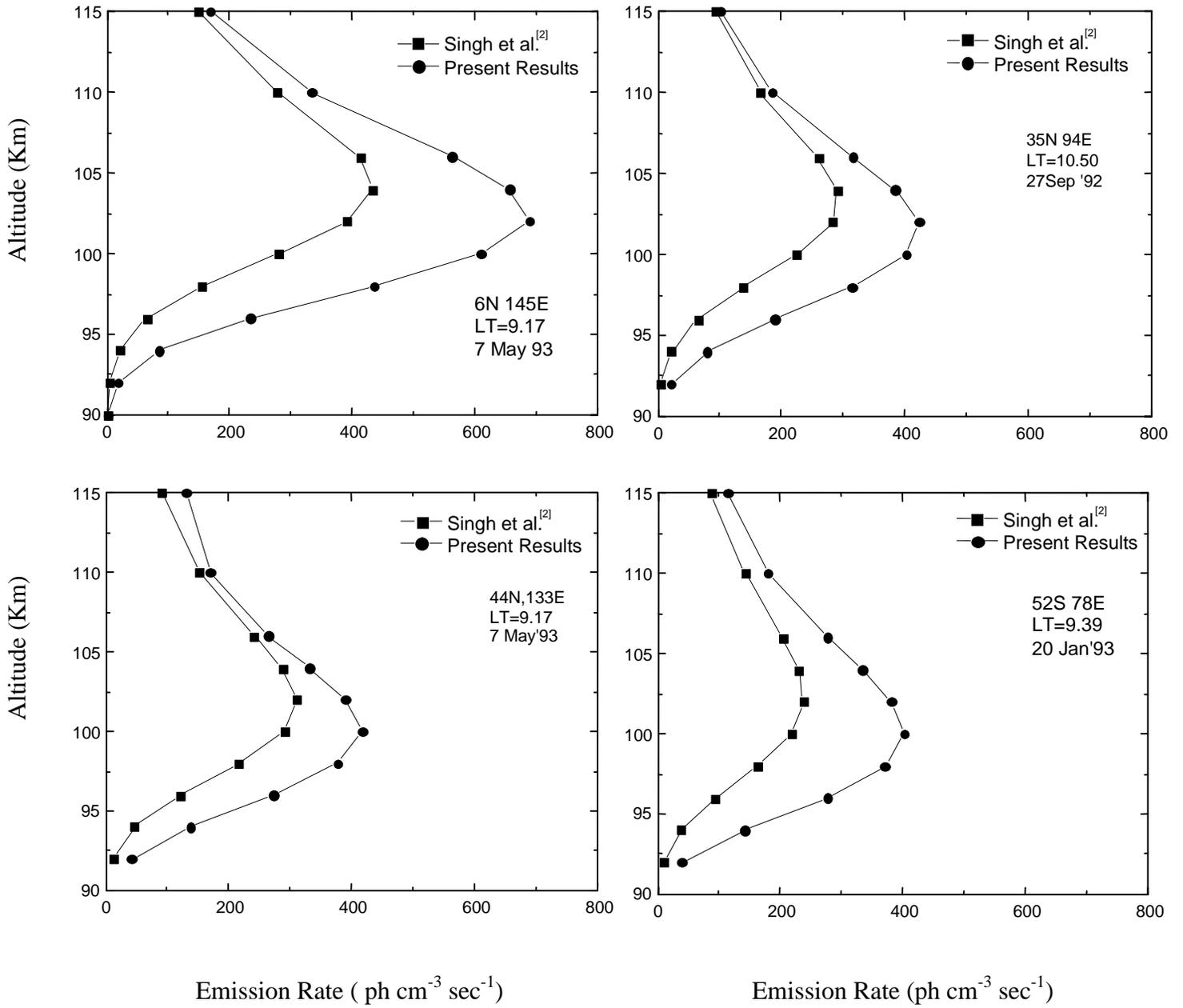

Figure 2

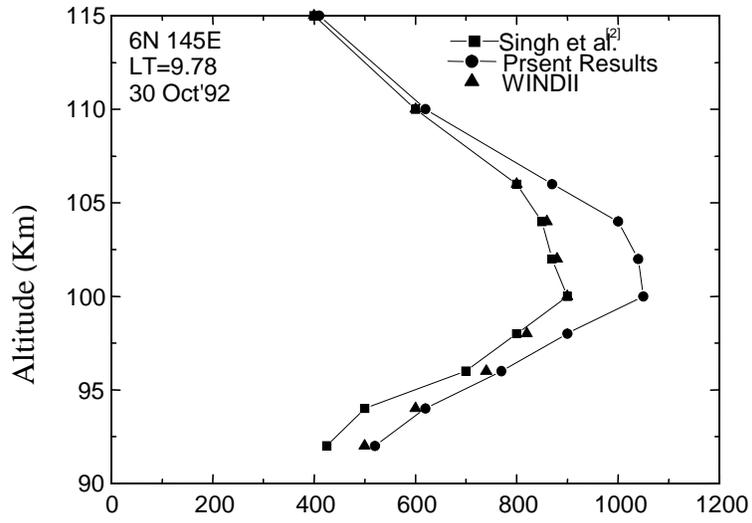

**(a)**

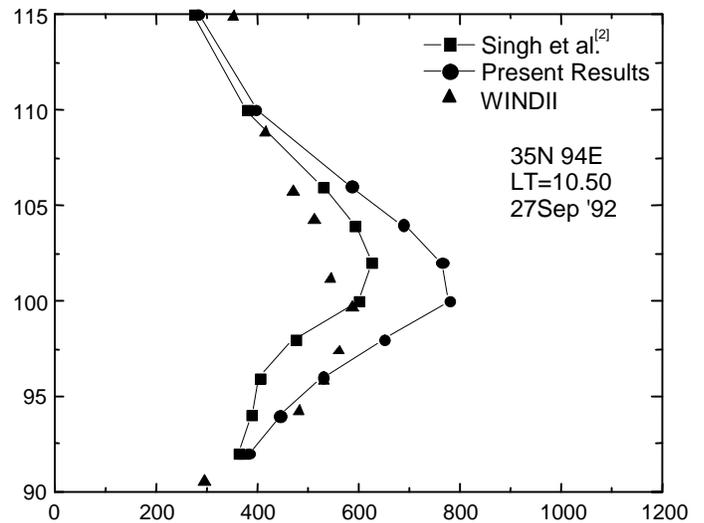

**(b)**

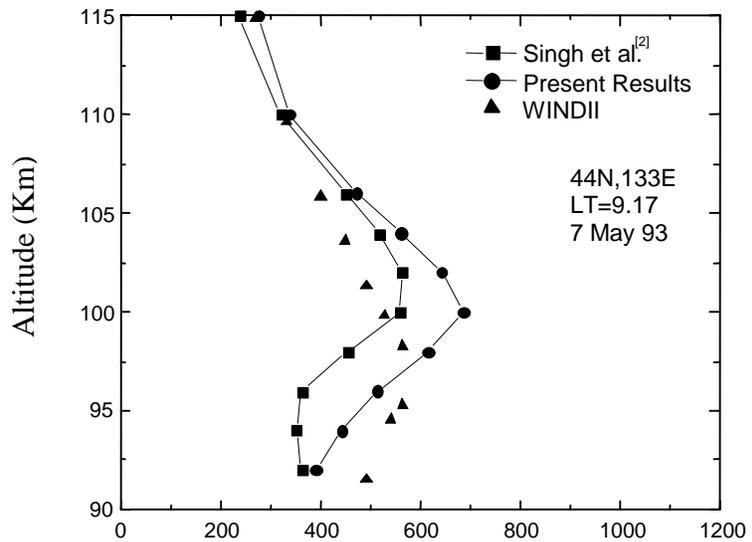

**(c)**

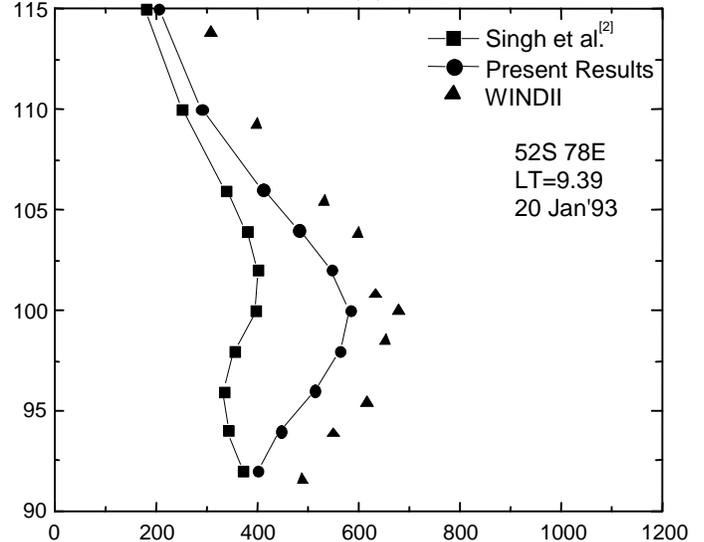

**(d)**

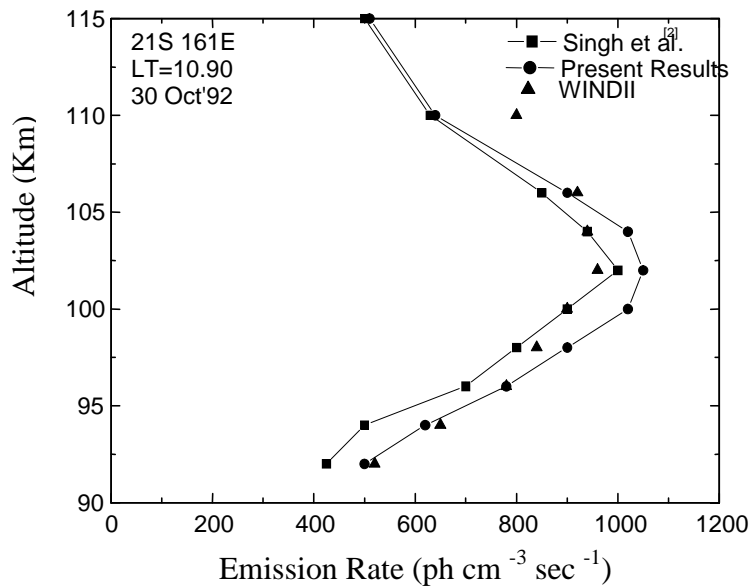

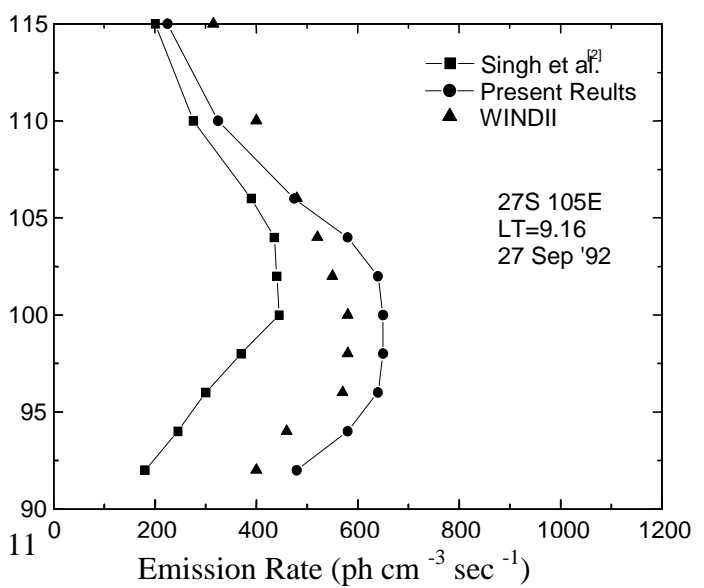

11